\documentclass[12pt]{article}

\usepackage[dvipsname]{xcolor}

\usepackage{graphicx}
\usepackage{amsmath}        
\usepackage{amssymb}

\def\mydate{April 23, 2019}
\def\ignore#1{{}}

\def\go{\rightarrow}

\def\eff{{\rm eff}}
\def\SM{{\rm SM}}
\def\KK{{\rm KK}}
\def\EM{{\rm EM}}

\def\la{\langle}
\def\ra{\rangle}

\def\mbig{\displaystyle }

\def\mfrac#1#2{\frac{\mbig #1}{\mbig #2}}

\makeatletter
\@addtoreset{equation}{section}
\makeatother

\begin{document}

\thispagestyle{empty}

{\small \noindent \mydate    \hfill OU-HET 1010}

\vskip 4.cm

\baselineskip=30pt plus 1pt minus 1pt

\begin{center}

{\Large \bf   Gauge-Higgs unification at $e^+ e^-$ linear colliders}

\end{center}


\baselineskip=22pt plus 1pt minus 1pt

\vskip 1.cm

\begin{center}
{\bf Yutaka Hosotani}

{\small \it Department of Physics, Osaka University, 
Toyonaka, Osaka 560-0043, Japan} \\

\end{center}

\vskip 2.cm
\baselineskip=18pt plus 1pt minus 1pt

\begin{abstract}
In gauge-Higgs unification the 4D Higgs boson appears as a part of the fifth dimensional component
of gauge potentials, namely as a fluctuation mode of the Aharonov-Bohm phase in the extra dimension.
The $SO(5) \times U(1) \times SU(3)$ gauge-Higgs unification gives nearly the same phenomenology as the 
standard model (SM) at low energies.  It predicts KK excited states of photon, $Z $ boson, and $Z_R$ 
boson ($Z'$ bosons) around 7 - 8 TeV. Quarks and leptons couple to these $Z'$ bosons with large 
parity violation, which leads to distinct interference effects in $e^+ e^-  \rightarrow \mu^+ \mu^-, q \, \bar q$
processes.  At 250 GeV ILC with polarized electron beams, deviation from SM can be seen 
at the 3 - 5 sigma level even with 250 fb$^{-1}$ data, namely in the early stage of ILC. 
Signals become stronger at higher energies. Precision measurements of interference effects at electron-positron colliders at energies above 250 GeV become very important to explore physics beyond the standard model.   
\end{abstract}


\vfill

\noindent
\hrulefill

\baselineskip=14pt 
{\footnotesize
To appear in the Proceedings of {\it Corfu Summer Institute 2018 
``School and Workshops on Elementary Particle Physics and Gravity"}  (CORFU2018),
31 August - 28 September, 2018, Corfu, Greece.}

\newpage
\baselineskip=20pt plus 1pt minus 1pt
\parskip=0pt

\section{Gauge-Higgs unification}

The existence of the Higgs boson of a mass  $125\,$GeV has been firmly confirmed. 
It  establishes the unification scenario of electromagnetic and weak forces.
In the standard model (SM) electromagnetic and weak forces are unified as $SU(2)_L \times U(1)_Y$
gauge forces.  The $SU(2)_L \times U(1)_Y$ gauge symmetry is spontaneously broken
by the Higgs scalar fields, whose neutral component appears as the observed Higgs boson.
Although almost all experimental data are consistent with the SM, it is not clear whether  
the observed Higgs boson is precisely what the SM assumes to exit.

The gauge sector of the SM is beautiful.  The gauge principle dictates how  quarks and leptons
interact with each other by gauge forces.  In the SM the Higgs field gives masses to 
quarks, leptons, and weak bosons.  However, the potential for the Higgs boson must be
prepared by hand such that it induces the spontaneous breaking of $SU(2)_L \times U(1)_Y$
symmetry.  
To put it differently, there is no principle for the Higgs field which determines how the Higgs
boson interacts with itself and other fields.  The lack of a principle results in the arbitrariness in
the choice of  parameters in the theory.
Furthermore the Higgs boson acquires an infinitely large correction to its mass at the quantum level
which must be cancelled by fine tuning of bare parameters.  It is called as the gauge hierarchy problem.
In addition,  even the ground state for the Higgs boson may become unstable against quantum corrections.

Gauge-Higgs unification (GHU) naturally solves those problems.
The 4d Higgs boson appears as a fluctuation mode of an Aharonov-Bohm (AB) phase in
the fifth dimension of spacetime, thus becoming a part of gauge fields.  
By dynamics of the AB phase the Higgs boson acquires a finite mass at the quantum level,
which is protected from divergence by the gauge principle.
The interactions of the Higgs boson are governed by the gauge principle, too.
In short, gauge fields and the Higgs boson are unified.\cite{Hosotani1983}-\cite{Hatanaka1998}

A realistic model of gauge-Higgs unification has been proposed.  
It is the $SO(5) \times U(1)_X \times SU(3)_C$ gauge-Higgs unification in the Randall-Sundrum 
warped space.  It  reproduces the SM content of gauge fields and matter content, and
SM phenomenology at low energies.
It leads to small deviations in the Higgs couplings.  It also predicts  new particles at the scale 5 TeV 
to 10 TeV as Kaluza-Klein (KK) excitation modes in the fifth dimensions.
Signals of these new particles can be seen both at LHC and at ILC.\cite{Kubo2002}-\cite{FHHOY2019}

One of the distinct features of the gauge-Higgs unification is large parity violation in the couplings
of quarks and leptons to KK exited states of gauge bosons.  
Right-handed quarks and leptons have much larger couplings to the first KK excited states of photon, 
$Z$ boson, and $Z_R$ boson (called as $Z'$ bosons) than the left-handed ones.
These $Z'$ bosons have masses around 7 TeV - 8 TeV.  
We will show below that even at 250 GeV ILC with 250 fb$^{-1}$ data
large deviations from the SM in various cross sections in $e^+ e^- \go f \bar f$ processes
can be seen by measuring the dependence on the polarization of the electron beam.
The key technique is to see  interference effects between the contribution from photon and $Z$ 
boson and the contribution from $Z'$ bosons.

We comment that there might be variation in the matter content of the 
$SO(5) \times U(1)_X \times SU(3)_C$ gauge-Higgs unification.  Recently a new way of 
introducing quark and lepton multiplets has been found, which can be embedded
in the $SO(11)$ gauge-Higgs grand unification.\cite{FHHOY2019}
Other options for fermion content have been proposed.\cite{Yoon2018}
These models can be clearly distinguished from each other
by investigating the polarization dependence of electron/positron beams in fermion pair production at ILC.
Note also that gauge-Higgs unification scenario provides  new approaches to dark matter, 
Higgs, and neutrino physics.\cite{FHHOS-DM2014}-\cite{Lim2018}

\section{$SO(5) \times U(1)\times SU(3)$ GHU in Randall-Sundrum warped space}

The theory is defined in the Randall-Sundrum (RS) warped space whose metric is given by
\begin{align}
ds^2= e^{-2\sigma(y)} \eta_{\mu\nu}dx^\mu dx^\nu+dy^2,
\label{RSmetric1}
\end{align}
where $\mu,\nu=0,1,2,3$,  $\eta_{\mu\nu}=\mbox{diag}(-1,+1,+1,+1)$,
$\sigma(y)=\sigma(y+ 2L)=\sigma(-y)$, and $\sigma(y)=ky$ for $0 \le y \le L$.
It has the topological structure $S^1/Z_2$.
In terms of the conformal coordinate $z=e^{ky}$
($1\leq z\leq z_L=e^{kL}$) in the region $0 \leq y \leq L$ 
\begin{align}
ds^2=  \frac{1}{z^2} \bigg(\eta_{\mu\nu}dx^{\mu} dx^{\nu} + \frac{dz^2}{k^2}\bigg) .
\label{RSmetric-2}
\end{align}
The bulk region $0<y<L$ ($1<z<z_L$) is anti-de Sitter (AdS) spacetime 
with a cosmological constant $\Lambda=-6k^2$, which is sandwiched by the
UV brane at $y=0$ ($z=1$) and the IR brane at $y=L$ ($z=z_L$).  
The KK mass scale is $m_{\rm KK}=\pi k/(z_L-1) \simeq \pi kz_L^{-1}$
for $z_L\gg 1$.

Gauge fields $A_M^{SO(5)}$, $A_M^{U(1)_X}$ and $A_M^{SU(3)_C}$ of 
$SO(5) \times U(1)_X \times SU(3)_C$, with gauge couplings
$g_A$, $g_B$ and $g_C$,  satisfy the orbifold conditions\cite{HOOS2008, FHHOS2013, FHHOY2019}
\begin{align}
&\begin{pmatrix} A_\mu \cr  A_{y} \end{pmatrix} (x,y_j-y) =
P_{j} \begin{pmatrix} A_\mu \cr  - A_{y} \end{pmatrix} (x,y_j+y)P_{j}^{-1}
\label{BC-gauge1}
\end{align}
where $(y_0, y_1) = (0, L)$.  For $A_M^{SO(5)}$
\begin{align}
P_0=P_1 = P_{\bf 5}^{SO(5)} = \mbox{diag}  (I_{4},-I_{1} ) ~,
\label{BC-matrix1}
\end{align}
whereas  $P_0=P_1=I$ for $A_M^{U(1)_X}$ and $A_M^{SU(3)_C}$.
With this set of boundary conditions $SO(5)$ gauge symmetry is broken to
$SO(4) \simeq SU(2)_L \times SU(2)_R$.  At this stage there appear zero modes
of 4D gauge fields in $SU(3)_C$, $SU(2)_L \times SU(2)_R$ and $U(1)_X$.
There appear zero modes in the $SO(5)/SO(4)$ part of $A_y^{SO(5)}$,
which constitute an $SU(2)_L$ doublet and become 4D Higgs fields.
As a part of gauge fields the 4D Higgs boson $H(x)$ appears as an AB phase in the fifth
dimension;
\begin{align}
&\hat W = P \exp \bigg\{ i g_A \int_{-L}^L dy \, A_y \bigg\}  \cdot P_1 P_0 
\sim \exp \bigg\{ i \bigg(\theta_H  + \frac{H(x)}{f_H} \bigg) 2 T^{(45)} \bigg\} ~,  
\label{ABphase1}
\end{align}
where
\begin{align}
&f_H = \frac{2}{g_w} \sqrt{ \frac{k}{L(z_L^2 -1)}} \sim 
\frac{2 ~ m_\KK}{\pi g_w \sqrt{kL}}  ~.
\label{ABphase2}
\end{align}
$g_w = g_A/\sqrt{L}$ is the 4D  weak coupling.
Gauge invariance implies that physics is periodic in $\theta_H $ with a period $2\pi$.

A brane scalar field $\hat \Phi_{(1,2,2, \frac{1}{2})} (x)$ or $\hat \Phi_{(1,1,2, \frac{1}{2})} (x)$ is
introduced on the UV brane where subscripts indicate the 
$SU(3)_C \times SU(2)_L \times SU(2)_R \times U(1)_X$ content.
Nonvanishing $\la \hat \Phi \ra$ spontaneously breaks $SU(2)_R \times U(1)_X$ 
to $U(1)_Y$, resulting in the SM symmetry $SU(3)_C \times SU(2)_L \times U(1)_Y$.

Once the fermion content is specified, the effective potential $V_\eff (\theta_H)$ is
evaluated.  The location of the global minimum of $V_\eff (\theta_H)$ determines
the value of $\theta_H$.  When $\theta_H \not= 0$, $SU(2)_L \times U(1)_Y$ symmetry
is dynamically broken to $U(1)_\EM$.  It is called the Hosotani mechanism.\cite{Hosotani1983}
The $W$ boson mass is given by
\begin{align}
m_W \sim \sqrt{\frac{k}{L}} ~ z_L^{-1} \, \sin \theta_H
\sim \frac{\sin \theta_H}{\pi \sqrt{kL}} ~ m_\KK ~.
\label{Wmass1}
\end{align}
As typical values, for $\theta_H = 0.10$ and $z_L = 3.6 \times 10^4$ one find
$m_\KK = 8.1\,$TeV and $f_H = 2.5\,$TeV.
There appears natural little hierarchy in the weak scale ($m_Z$) and the KK scale ($m_\KK$).

Quark and lepton multiplets are introduced in the vector representation {\bf 5} of $SO(5)$.
Further dark fermions are introduced in the spinor representation {\bf 4} of $SO(5)$.
This model is called as the A-model, and has been investigated intensively 
so far.\cite{FHHOS2013}-\cite{FHHO2017ILC}
Recently  an alternative way of introducing matter has been found.\cite{FHHOY2019}   
This model, called as the B-model, can be implemented in the $SO(11)$ gauge-Higgs 
grand unification.\cite{HosotaniYamatsu2015, Furui2016,  HosotaniYamatsu2017}
The matter content of the two models is summarized in Table \ref{Table-matter}.
In this talk phenomenological consequences of the A-model are presented.

\begin{table}[tbh]
{
\renewcommand{\arraystretch}{1.4}
\begin{center}
\caption{Matter fields.   $SU(3)_C\times SO(5) \times U(1)_X$ content
is shown.   In the  A-model only $SU(3)_C\times SO(4) \times U(1)_X$
symmetry is maintained on the UV brane so that the $SU(2)_L \times SU(2)_R$ content
is shown for brane fields.  In the B-model  given in ref.\ \cite{FHHOY2019}  the full 
$SU(3)_C\times SO(5) \times U(1)_X$ invariance is preserved on the UV brane.
}
\vskip 10pt
\begin{tabular}{|c|c|c|}
\hline
&A-model 
&B-model 
\\
\hline \hline
quark
&$({\bf 3}, {\bf 5})_{\frac{2}{3}} ~ ({\bf 3}, {\bf 5})_{-\frac{1}{3}}$ 
&$({\bf 3}, {\bf 4})_{\frac{1}{6}} ~ ({\bf 3}, {\bf 1})_{-\frac{1}{3}}^+ 
    ~ ({\bf 3}, {\bf 1})_{-\frac{1}{3}}^-$
\\
\hline
lepton
&$({\bf 1}, {\bf 5})_{0} ~ ({\bf 1}, {\bf 5})_{-1}$  
&$\strut ({\bf 1}, {\bf 4})_{-\frac{1}{2}}$ 
\\
\hline
dark fermion 
&$({\bf 1}, {\bf 4})_{\frac{1}{2}}$ 
& $({\bf 3}, {\bf 4})_{\frac{1}{6}} ~ ({\bf 1}, {\bf 5})_{0}^+ ~ ({\bf 1}, {\bf 5})_{0}^-$ 
\\
\hline \hline
brane fermion 
&$\begin{matrix} ({\bf 3}, [{\bf 2, 1}])_{\frac{7}{6}, \frac{1}{6}, -\frac{5}{6}} \cr
\noalign{\kern -4pt}
({\bf 1}, [{\bf 2, 1}])_{\frac{1}{2}, -\frac{1}{2}, -\frac{3}{2}} \end{matrix}$
&$({\bf 1}, {\bf 1})_{0} $ 
\\
\hline
brane scalar 
&$({\bf 1}, [{\bf 1,2}])_{\frac{1}{2}}$ 
&$({\bf 1}, {\bf 4})_{\frac{1}{2}} $ 
\\
\hline
Sym.\ on UV brane
&$SU(3)_C \times SO(4) \times U(1)_X$ 
&$SU(3)_C \times SO(5) \times U(1)_X$ 
\\
\hline
\end{tabular}
\label{Table-matter}
\end{center}
}
\end{table}

The correspondence between the SM in four dimensions and the gauge-Higgs unification
in five dimensions is summarized as
\begin{align}
\begin{matrix}
{\rm SM} && {\rm GHU} \cr
\noalign{\kern 3pt}
\displaystyle \int d^4x \Big\{ {\cal L}^{\rm gauge} + {\cal L}^{\rm Higgs}_{\rm kinetic} \Big\}
&~ \Rightarrow ~
&\displaystyle \int d^5 x \sqrt{-g} ~ {\cal L}^{\rm gauge}_{\rm 5d} \cr
\noalign{\kern 5pt}
\displaystyle \int d^4 x \Big\{ {\cal L}^{\rm fermion} + {\cal L}^{\rm Yukawa} \Big\}
&\Rightarrow
&\displaystyle \int d^5 x \sqrt{-g} ~ {\cal L}^{\rm fermion}_{\rm 5d} \cr
\noalign{\kern 5pt}
- \displaystyle \int d^4 x ~ {\cal L}^{\rm Higgs}_{\rm potential}
&\Rightarrow
& \displaystyle \int d^4 x   ~ V_\eff (\theta_H)
\end{matrix}
\label{correspondence}
\end{align}
In the SM, ${\cal L}^{\rm gauge}$, $ {\cal L}^{\rm Higgs}_{\rm kinetic}$ and ${\cal L}^{\rm fermion}$
are governed by the gauge principle, but ${\cal L}^{\rm Yukawa}$ and ${\cal L}^{\rm Higgs}_{\rm potential}$
are not.  On the GHU side in (\ref{correspondence}), ${\cal L}^{\rm gauge}_{\rm 5d}$ and 
${\cal L}^{\rm fermion}_{\rm 5d}$ are governed by the gauge principle and 
$V_\eff (\theta_H)$ 
follows from them.

\section{Gauge couplings and Higgs couplings}
\begin{figure}[tbh]
\begin{center}
\includegraphics[bb=0 0 360 255, height=4.5cm]{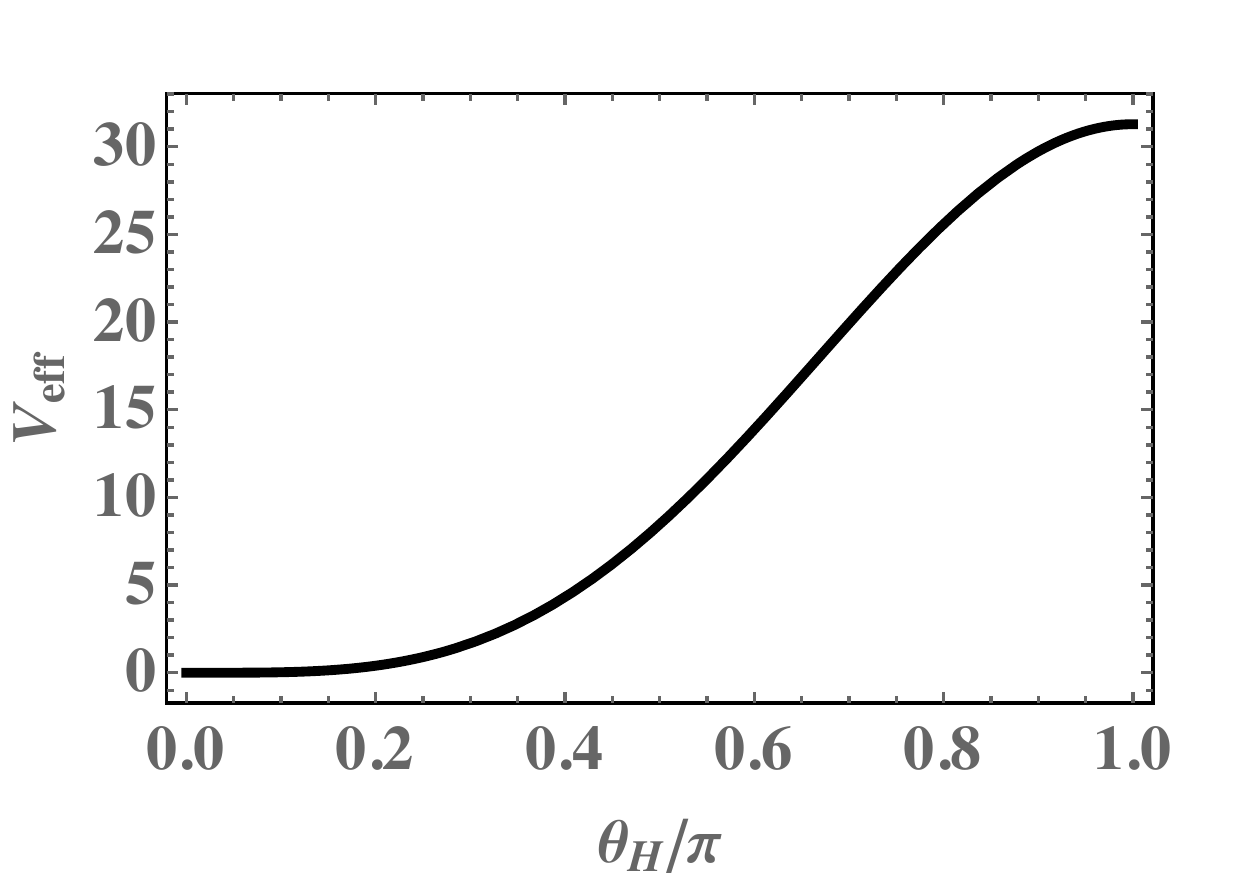}
\includegraphics[bb=0 0 360 227, height=4.5cm]{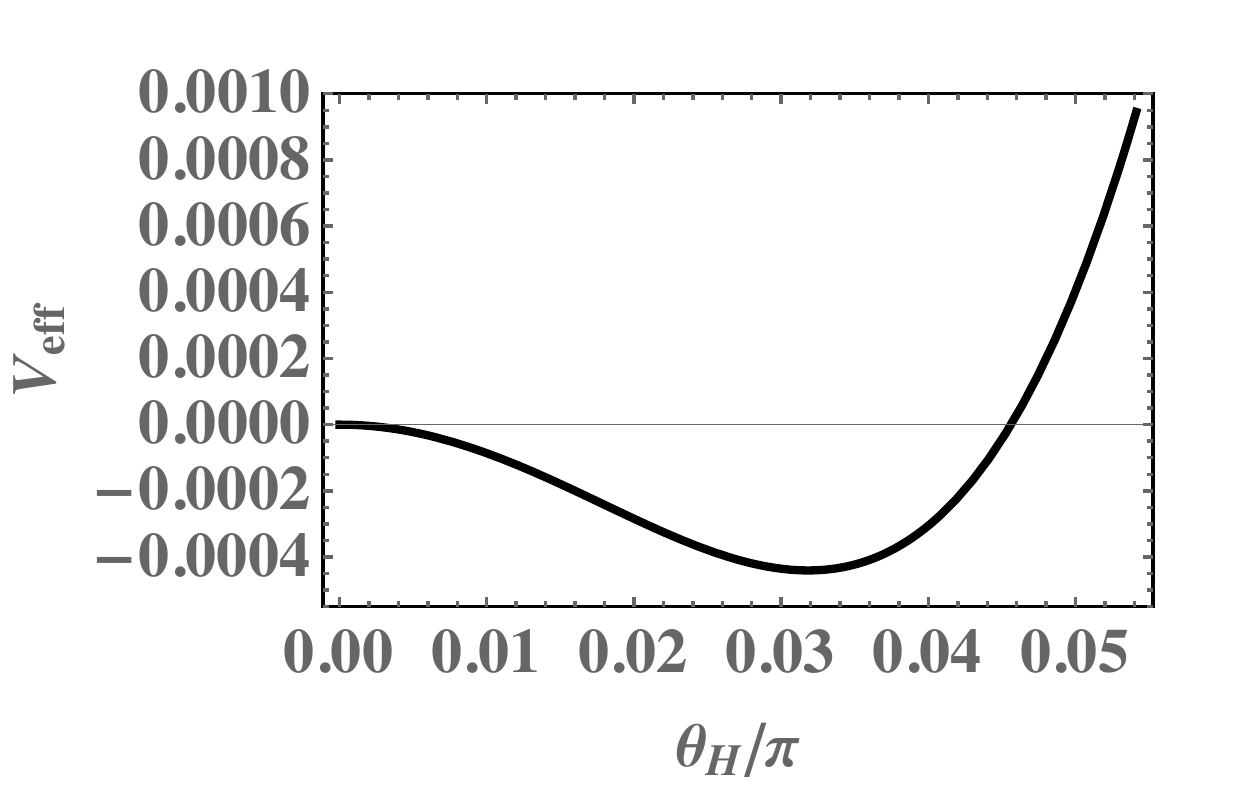}
\end{center}
\vskip -10pt
\caption{Effective potential $V_\eff (\theta_H)$ is displayed in the unit of  $(k z_L^{-1})^4/16 \pi^2$
for $z_L = 3.56 \times 10^4$ and $m_\KK = 8.144\,$TeV.
The minimum of $V_\eff$ is located at $\theta_H = 0.10$. 
The curvature at the minimum determines the Higgs boson mass by
$m_H^2 = f_H^{-2} V_\eff''(\theta_H)|_{\rm min}$, yielding  $m_H = 125.1\,$GeV. }
\label{figure-Veff}
\end{figure}

Let us focus on the A-model.
The SM quark-lepton content is reproduced with no  exotic light fermions.  
The  one-loop effective potential $V_\eff (\theta_H)$ is displayed by in fig.\ \ref{figure-Veff}.
The finite Higgs boson mass $m_H \sim 125\,$GeV is generated  naturally  with $\theta_H \sim 0.1$.
Relevant parameters in the theory are determined from quark-lepton masses, 
$m_Z$, and electromagnetic, weak, and strong gauge coupling constants.
Many of physical quantities depend on the value of $\theta_H$, but not on other parameters.
In the SM the $W$ and $Z$ couplings of quarks and leptons are universal.
They depend on only representations of the group $ SU(2)_L \times U(1)_Y$.
In GHU the $W$ and $Z$ couplings of quarks and leptons may depend on more detailed
behavior of wave functions in the fifth dimension.
Four-dimensional couplings are obtained by integrating the product of the 
$W/Z$ and quark/lepton wave functions over the fifth dimensional coordinate.

Surprisingly the $W$ and $Z$ couplings of quarks and leptons and the $WWZ$ coupling
in GHU turn out very close to those in the SM.  
The result is tabulated in Table \ref{Table-gaugeCoupling1}.
In the last column the values in the SM are listed.
The deviations from the SM are very small.
The $W$ couplings of left-handed light quarks and leptons are approximately
given by
\begin{align}
g_L^W\sim  g_w \, \frac{\sqrt{2kL}}{\sqrt{2kL - \frac{3}{4} \sin^2 \theta_H }}
\sim g_w \, \Big( 1 + \frac{3 \sin^2 \theta_H}{16 kL} \Big) ~.
\label{Wcoupling1}
\end{align}
Here $kL = \ln z_L$.   The $W$ couplings of right-handed quarks and leptons are
negligibly small.

\begin{table}[tbh]
\renewcommand{\arraystretch}{1.1}
\begin{center}
\caption{Gauge ($W, Z$) couplings of quarks and leptons. $WWZ$ coupling is also listed 
at the bottom.  The values in the SM are listed in the last column.
}
\vskip 10pt
\label{Table-gaugeCoupling1}
\begin{tabular}{|c|c|cc|cc|c|}
\hline
\multicolumn{2}{|c|}{~}
&\multicolumn{2}{|c|}{$\theta_H = 0.115$}
 &\multicolumn{2}{|c|}{$\theta_H=0.0737$} & SM\\
\hline
&$(\nu_e , e)$ 
&\multicolumn{2}{|c|}{1.00019} &\multicolumn{2}{|c|}{ 1.00009} & \\
&$(\nu_\mu , \mu)$ 
&\multicolumn{2}{|c|}{1.00019} & \multicolumn{2}{|c|}{1.00009 }&1 \\
$g_L^W/g_w$
&$(\nu_\tau , \tau)$ &\multicolumn{2}{|c|}{1.00019} & \multicolumn{2}{|c|}{1.00009 } & \\
\cline{2-7}
&$(u,d)$ &\multicolumn{2}{|c|}{1.00019} & \multicolumn{2}{|c|}{1.00009 } & \\
&$(c,s)$ &\multicolumn{2}{|c|}{1.00019} & \multicolumn{2}{|c|}{1.00009 } &1 \\
&$(t,b)$ &\multicolumn{2}{|c|}{0.9993} & \multicolumn{2}{|c|}{0.9995} & \\
\hline
&$\nu_e, \nu_\mu, \nu_\tau$ 
&0.50014 &0 &0.50008 &0  & 0.5 \qquad  0\\
\cline{2-7}
 &$e, \mu, \tau$ 
&-0.2688 &0.2314 &-0.2688 &0.2313  & -0.2688 $\,$ 0.2312\\
\cline{2-7}
$(g_L^Z,g_R^Z)/g_w$   &$u, c$ 
&0.3459 &-0.1543 &0.3459 &-0.1542  & 0.3458 $\,$ -0.1541\\
&$t$ 
&0.3449 &-0.1553 &0.3453 &-0.1549  & \\
\cline{2-7}
&$d, s$ 
&-0.4230 &0.0771 &-0.4230 &0.0771  & -0.4229 $\,$ 0.0771\\
&$b$ 
&-0.4231 &0.0771 &-0.4230 &0.0771  & \\
\hline
\multicolumn{2}{|c|}{$g_{WWZ}/g_w \cos \theta_W$}
&\multicolumn{2}{|c|}{0.9999998} &\multicolumn{2}{|c|}{0.99999995} &1\\
\hline
\end{tabular}
\end{center}
\end{table}

\def\mynoalign1{\noalign{\kern 3pt}}

Yukawa couplings of quarks and leptons, and $WWH$, $ZZH$ couplings are well approximated by
\begin{align}
\begin{pmatrix} g_{\rm Yukawa} \cr \mynoalign1 g_{WWH}   \cr \mynoalign1 g_{ZZH}  \end{pmatrix}
&\sim 
\begin{pmatrix} g_{\rm Yukawa}^\SM \cr  \mynoalign1 g_{WWH}^\SM \cr 
\mynoalign1 g_{ZZH}^\SM \end{pmatrix} \times \cos\theta_H 
\label{HiggsCoupling1}
\end{align}
where $g_{\rm Yukawa}^\SM$ on the right side, for instance,  denotes the value in the SM.
For $\theta_H \sim 0.1$ the deviation amounts to only 0.5\%.
Larger deviations are expected in the cubic and quartic self-couplings of the Higgs boson.
They are approximately given by
\begin{align}
\lambda_3^{\rm Higgs} &\sim 156.9 \, \cos\theta_H + 17.6 \, \cos^2 \theta_H ~~({\rm GeV}), \cr
\noalign{\kern 5pt}
\lambda_4^{\rm Higgs} &\sim - 0.257+ 0.723\cos 2 \theta_H + 0.040 \cos 4 \theta_H ~.
\label{HiggsCoupling2}
\end{align}
In the SM, $\lambda_3^{\rm Higgs, SM} = 190.7\,$GeV and $\lambda_4^{\rm Higgs, SM} = 0.774$.
In the $\theta_H \go 0$ limit $\lambda_3^{\rm Higgs}$ and $\lambda_4^{\rm Higgs}$ become
8.5\% and 35\% smaller than the values in the SM.    
$\lambda_3^{\rm Higgs}$ can be measured at ILC.

GHU gives nearly the same phenomenology at low energies as the SM.
To  distinguish GHU from the SM, one need to look at signals of new particles which GHU predicts.

\section{New particles -- KK excitation}

KK excitations of each particle appear as new particles.  The existence of an extra dimension
is confirmed by observing  KK excited particles of quarks, leptons, and gauge bosons.
The KK spectrum is shown in Table \ref{table-KKspectrum1}.
$Z_R$ is the gauge field associated with $SU(2)_R$, and has no zero mode.
$Z^{(1)}$, $\gamma^{(1)}$ and $Z_R^{(1)}$ are called as $Z'$ bosons.
Clean signals can be found in the process $q \, \bar q \go Z' \go e^+ e^- , \mu^+ \mu^-$
at LHC.  So far no event of $Z'$ has been observed, which puts the limit $\theta_H < 0.11$.

\begin{table}[htb]
\caption{The mass spectrum $\{ m_n \}$ ($n \ge 1$) of KK excited modes of gauge bosons and  quarks
for $\theta_H = 0.10, n_F = 4$, where $n_F$ is the number of dark fermion multiplets.
Masses are given in the unit of TeV.
Pairs $(W^{(n)}, Z^{(n)})$, $(W_R^{(n)}, Z_R^{(n)})$, $(t^{(n)}, b^{(n)})$, $(c^{(n)}, s^{(n)})$,
$(u^{(n)}, d^{(n)})$ ($n \ge 1$) have almost degenerate masses.
The spectrum of $W_R$ tower is the same as that of $Z_R$ tower.
The gluon tower has the same spectrum as the photon ($\gamma$) tower.
}
\label{table-KKspectrum1}
\vskip 5pt
\centering
\renewcommand{\arraystretch}{1.1}
\begin{tabular}{|c|c|c|c|c||c|c|c|c|}
\hline
\multicolumn{9}{|c|}{$\theta_H = 0.10, ~ n_F = 4, ~ m_\KK = 8.144 \,{\rm TeV}, ~ z_L = 3.56 \times 10^4$}     \\
\hline
&\multicolumn{2}{c|}{$Z^{(n)}$} &$\gamma^{(\ell)}$ &$Z_R^{(\ell)}$ 
&\multicolumn{2}{c|}{$t^{(n)}$}   &$c^{(n)}$  &$u^{(n)}$     \\
\hline
$n\,  (\ell)$ &$m_n$ &$\mfrac{m_n}{m_\KK}$ &$m_\ell$ &$m_\ell$ 
&$m_n$  &$\mfrac{m_n}{m_\KK}$ &$m_n$ &$m_n$  \\
\hline
$1\, (1)$&6.642 &$0.816$ &6.644 &6.234 
&7.462&0.916 &8.536 &10.47  \\
2 ~~~~~&9.935 &1.220 &-- &-- 
&8.814  &1.082 &12.01 &13.82  \\
$3\,  (2)$&14.76 &1.812&14.76 &14.31
&15.58 &1.913 &16.70 &18.76  \\
4 ~~~~~&18.19 &2.233 &-- &-- 
&16.99  &2.087 &20.41 &22.37  \\
\hline
\end{tabular}
\end{table}

The KK mass scale  as a function of $\theta_H$ is approximately given by
\begin{align}
&m_\KK (\theta_H) \sim \frac{1.36\,{\rm TeV}}{(\sin \theta_H )^{0.778}} ~, 
\label{KKscale1}
\end{align}
irrespective of the other parameters of the theory.  
In GHU many of physical quantities such as the Higgs couplings in (\ref{HiggsCoupling1}) 
and (\ref{HiggsCoupling2}), the KK scale (\ref{KKscale1}),  and KK masses of gauge bosons
are approximately determined by the value of $\theta_H$ only.  
This property is called as the $\theta_H$ universality.
Once the $Z^{(1)}$ particle  is found and its mass is determined, then the value of $\theta_H$
is fixed and the values of other physical quantities are predicted.\cite{FHHOS2013}

\begin{figure}[htb]
\begin{center}
\includegraphics[bb=10 57 690 191, width=10cm]{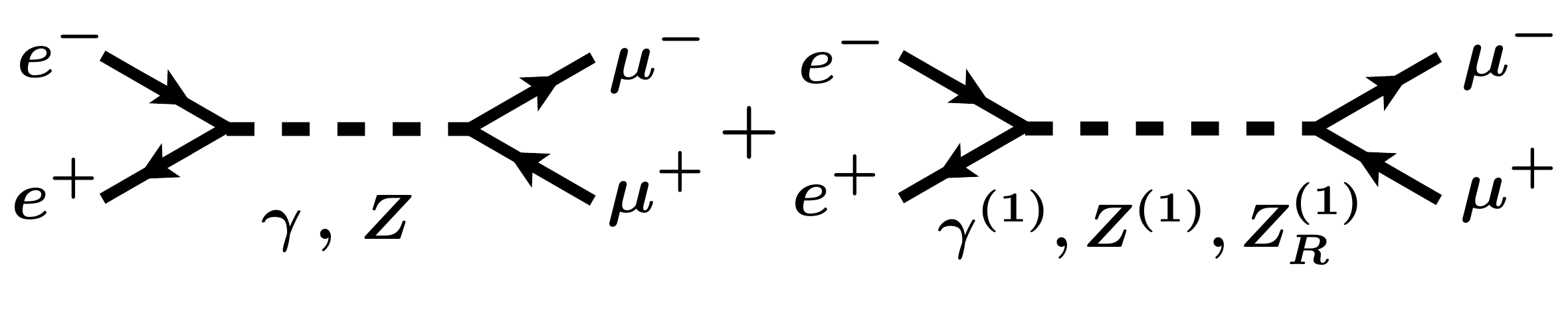}
\end{center}
\caption{
Dominant diagrams in the process
$e^+ e^- \go \mu^+ \mu^-$
}
\label{fig:ILC-Zprime}
\end{figure}

Although $Z'$ bosons are heavy with masses around 6 -- 8 TeV,  their effects can be
seen at 250 GeV ILC ($e^+ e^-$ collisions). (Fig.~\ref{fig:ILC-Zprime})
The couplings of right-handed quarks and leptons to $Z'$ bosons are much stronger 
than those of left-handed quarks and leptons.  This large parity violation manifests
as  an interference effect in $e^+ e^-$ collisions.\cite{FHHO2017ILC}

\begin{figure}[bht]
\begin{center}
\includegraphics[bb=2 88 425 375, width=8.0cm]{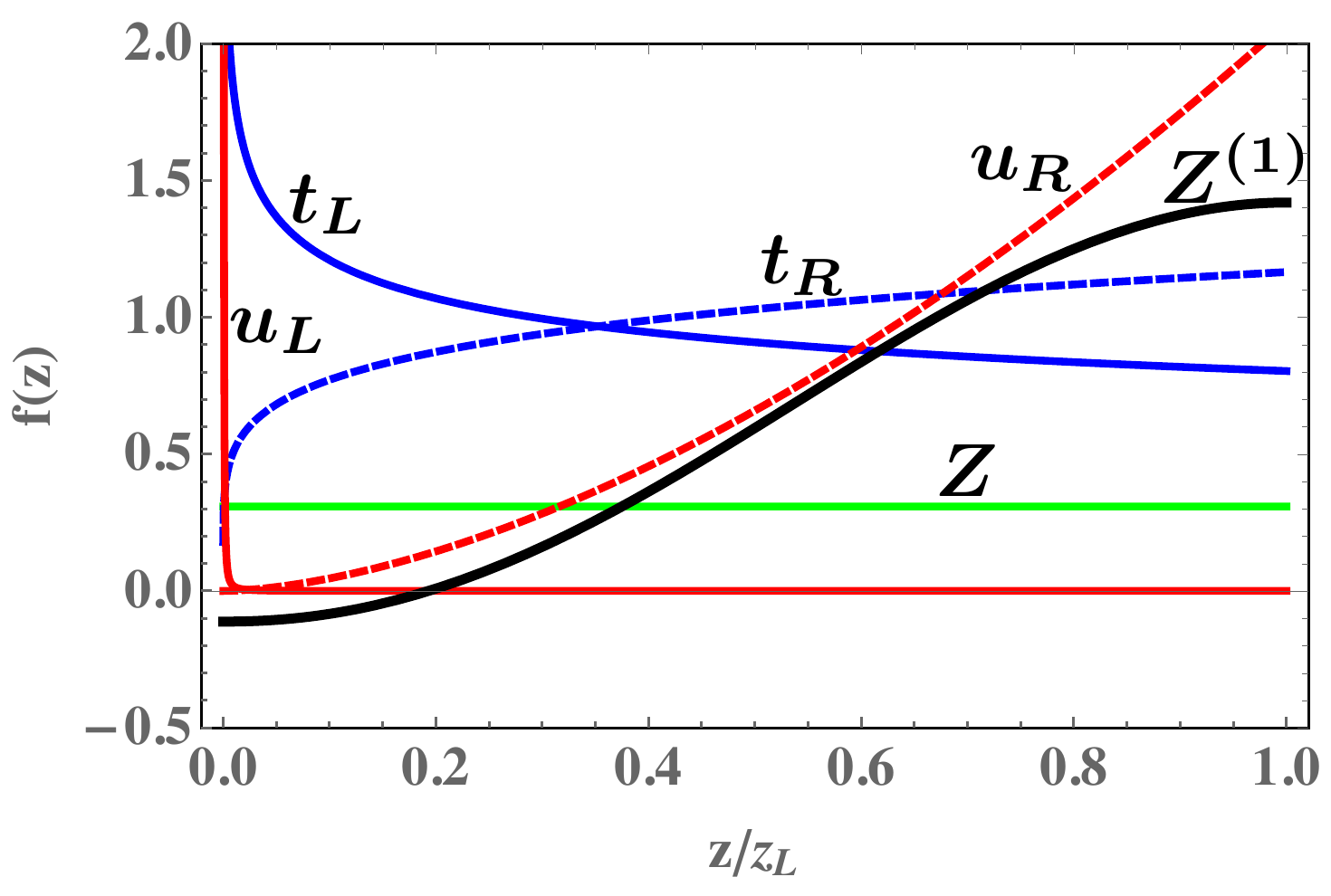}
\end{center}
\vskip  5pt
\caption{
Wave functions of various fermions and gauge bosons for $\theta_H = 0.1$.
Only some of the relevant components in $SO(5)$ are displayed.
Wave functions of light quarks and leptons are qualitatively similar to those of $(u_L, u_R)$.
Wave functions of $(b_L, b_R)$ are similar to those of $(t_L, t_R)$.
$Z$ boson wave function is almost constant, whereas 
$Z^{(1)}$'s wave function becomes large near the IR brane at $z=z_L$.
}
\label{fig-wavefunctions}
\end{figure}

Left-handed light quarks and leptons are localized near the UV brane (at $y=0$), whereas right-handed
ones near the IR brane (at $y=L$).  Wave functions of top and bottom quarks spread over the entire
fifth dimension.  In GHU both left- and right-handed fermions are in the same gauge multiplet so that
if a left-handed fermion is localized near the UV brane, then its partner right-handed fermion is 
necessarily localized near the IR brane.  KK modes of gauge bosons in the RS space are always
localized near the IR brane.   $Z'$ couplings of quarks and leptons are given by 
overlap-integrals of wave functions of $Z'$ bosons and left- or right-handed quarks and leptons.
Consequently right-handed quarks/leptons have larger couplings to $Z'$.
Typical behavior of wave functions is depicted in fig.~\ref{fig-wavefunctions}.

Gauge couplings of quarks and leptons to $Z^{(1)}$, $\gamma^{(1)}$ and $Z_R^{(1)}$ are 
summarized in Table~\ref{table-Zprimecoupling1}.
Except for $b$ and $t$ quarks, right-handed quarks and leptons have much larger couplings
than left-handed ones.

\begin{table}[h]
\caption{
Gauge couplings of quarks and leptons to $Z^{(1)}$, $\gamma^{(1)}$ and $Z_R^{(1)}$
for $\theta_H = 0.0917$ and $\sin^2 \theta_W = 0.2312$.
Couplings are given in the unit of $g_w/\cos \theta_W$.
The $Z$ couplings in the SM , $I_3 - \sin^2 \theta_W Q_\EM$, are also shown.
}
\label{table-Zprimecoupling1}
\vskip 5pt
\begin{center}
\renewcommand{\arraystretch}{1.0}
\begin{tabular}{|c|cc|cc|cc|cc|}
\hline
&\multicolumn{2}{c|}{SM: $Z$} 
&\multicolumn{2}{c|}{$Z^{(1)}$}
&\multicolumn{2}{c|}{$Z_R^{(1)}$}
&\multicolumn{2}{c|}{$\gamma^{(1)}$}\\
&Left &Right & Left &  Right & Left &  Right & Left &  Right  \\
\hline
$\nu_e$      &  &  & $-0.183$ & 0 & 0 & 0 & 0 & 0 \\
$\nu_{\mu}$  & $0.5$ & 0 & $-0.183$ & 0 & 0 & 0 & 0 & 0 \\
$\nu_{\tau}$ & $$ &  & $-0.183$ & 0 & 0 & 0 & 0 & 0 \\
\hline
$e$    
& $$ & $$ & $0.099$ & $0.916$ & 0 & $-1.261$ & $0.155$ & $-1.665$ \\
$\mu$  
&$-0.269$ &$0.231$ & $0.099$ & $0.860$ & 0 & $-1.193$ & $0.155$ & $-1.563$ \\
$\tau$ 
& $$ & $$ & $0.099$ & $0.814$ & 0 & $-1.136$ & $0.155$ & $-1.479$ \\
\hline
$u$ 
& $$ & $$ & $-0.127$ & $-0.600$ & $0$ & $0.828$ & $-0.103$ & $1.090$ \\
$c$ 
&$0.346$ &$-0.154$ & $-0.130$ & $-0.555$ & $0$ & $0.773$ & $-0.103$ & $1.009$ \\
$t$ 
& $$ & $$ & $0.494$ & $-0.372$ & $0.985$ & $0.549$& $0.404$ & $0.678$ \\
\hline
$d$ 
& $$ & $$ & $0.155$ & $0.300$ & $0$ & $-0.414$ & $0.052$ & $-0.545$ \\
$s$ 
&$-0.423$ &$0.077$ & $0.155$ & $0.277$ & $0$ & $-0.387$ & $0.052$ & $-0.504$ \\
$b$ 
& $$ & $$ & $-0.610$ & $0.186$ & $0.984$ & $-0.274$ & $-0.202$ & $-0.339$ \\
\hline
\end{tabular}
\end{center}
\end{table}

\section{$e^+ e^- $ collisions}

The amplitude ${\cal M}$  for  the $e^+ e^- \go \mu^+ \mu^-$ process at the tree level 
in  fig.~\ref{fig:ILC-Zprime}  can be expressed 
as the sum of two terms ${\cal M}_0$ and ${\cal M}_{Z'}$. 
\begin{align}
{\cal M} &= {\cal M}_0 + {\cal M}_{Z'} \cr
&= {\cal M}(e^+ e^- \go \gamma \, , \, Z \go \mu^+ \mu^-) 
+ {\cal M}(e^+ e^- \go  Z' \go \mu^+ \mu^-) ~.
\label{pairproduction1}
\end{align}
For $s = (250\,{\rm GeV})^2 \sim (1\,{\rm TeV})^2$, we have $m_Z^2 \ll s \ll m_{Z'}^2$
so that the amplitude can be approximated by  
\begin{align}
{\cal M} &\simeq \frac{g_w^2}{\cos^2 \theta_W}  \sum_{\alpha, \beta= L,R} J_\alpha^{(e)\nu} (p,p')
\bigg\{ \frac{\kappa_\SM^{\alpha\beta}}{s} - \frac{\kappa_{Z'}^{\alpha\beta}}{m_{Z'}^2} \bigg\}
J_{\beta\nu}^{(\mu)} (k,k')
\label{pairproduction2}
\end{align}
where $J_{\alpha\nu}^{(e)} (p,p')$ and  $J_{\beta\nu}^{(\mu)} (k,k')$ represent momentum and polarization
configurations of the initial and final states, respectively.
$\kappa_\SM^{\alpha\beta}$ and $\kappa_{Z'}^{\alpha\beta}$ are found from 
Table~\ref{table-Zprimecoupling1} to be
\begin{align}
(\kappa_\SM^{LL}, \kappa_\SM^{LR}, \kappa_\SM^{RL}, \kappa_\SM^{RR})
&= (0.25, 0.1156, 0.1156, 0.2312)~, \cr
\noalign{\kern 5pt}
(\kappa_{Z'}^{LL}, \kappa_{Z'}^{LR}, \kappa_{Z'}^{RL}, \kappa_{Z'}^{RR})~
&= (0.034, -0.158, -0.168, 4.895) ~.
\label{pairproduction3}
\end{align}
Compared with the value in the SM,  $\kappa_{Z'}^{RR}$ is very large whereas $\kappa_{Z'}^{LL}$
is very small.

Although direct production of $Z'$ particles is not possible with 
$s= (250\,{\rm GeV})^2 \sim (1\,{\rm TeV})^2$,  the interference term becomes appreciable.
Suppose that the electron beam is polarized in the right-handed mode.  Then the interference 
term gives
\begin{align}
\frac{{\cal M}_0  {\cal M}_{Z'}^*}{| {\cal M}_0 |^2} 
&\sim - \frac{ \kappa_{Z'}^{RR} +  \kappa_{Z'}^{RL}}{ \kappa_\SM^{RR} +  \kappa_\SM^{RL}} \,
\frac{s}{m_{Z'}^2} \sim - 13.6 \, \frac{s}{m_{Z'}^2} \cr
\noalign{\kern 10pt}
&\sim -0.017 \quad {\rm at} ~ \sqrt{s} = 250\,{\rm GeV} ~.
\label{pairproduction4}
\end{align}
This is a sufficiently big number.  As the number of events of fermion pair production is huge
in the proposed ILC experiment, 1.7\% correction can be certainly confirmed.
One recognizes that polarized electron and/or positron beams play an important role
to investigate physics beyond 
the SM.\cite{FHHO2017ILC}, \cite{Yoon2018}, \cite{Bilokin2017}-\cite{ILC2019}

\subsection{Energy and polarization dependence}

In the $e^+ e^-$ collision experiments one can control both the energy and polarization of 
incident electron and positron beams.  First consider  the total cross section for
$ e^+ e^- \go \mu^+ \mu^-$;
\begin{align}
F_1 = \frac{\sigma (e^+ e^- \go \mu^+ \mu^- )^{\rm GHU}}{\sigma (e^+ e^- \go \mu^+ \mu^- )^\SM} ~.
\label{mupair1}
\end{align}
Both the electron and positron beams are polarized with polarization $P_{e^-}$ and $P_{e^+}$.
For purely right-handed (left-handed) electrons $P_{e^-} = +1 (-1)$.
At $\sqrt{s} \ge 250\,$GeV, $e^+$ and  $e^-$ in the initial state may be viewed as massless particles.
The ratio $F_1$ in (\ref{mupair1}) depends on the effective polarization
\begin{align}
P_\eff = \frac{P_{e^-} - P_{e^+}}{1 - P_{e^-} P_{e^+} } ~.
\label{Peff1}
\end{align}
At the proposed 250 GeV ILC,  $|P_{e^-}| \le 0.8$ and $|P_{e^+}| \le 0.3$
so that $|P_\eff| \le 0.877$.

\begin{figure}[thb]
\begin{center}
\includegraphics[bb=0 0 288 293, width=6.0cm]{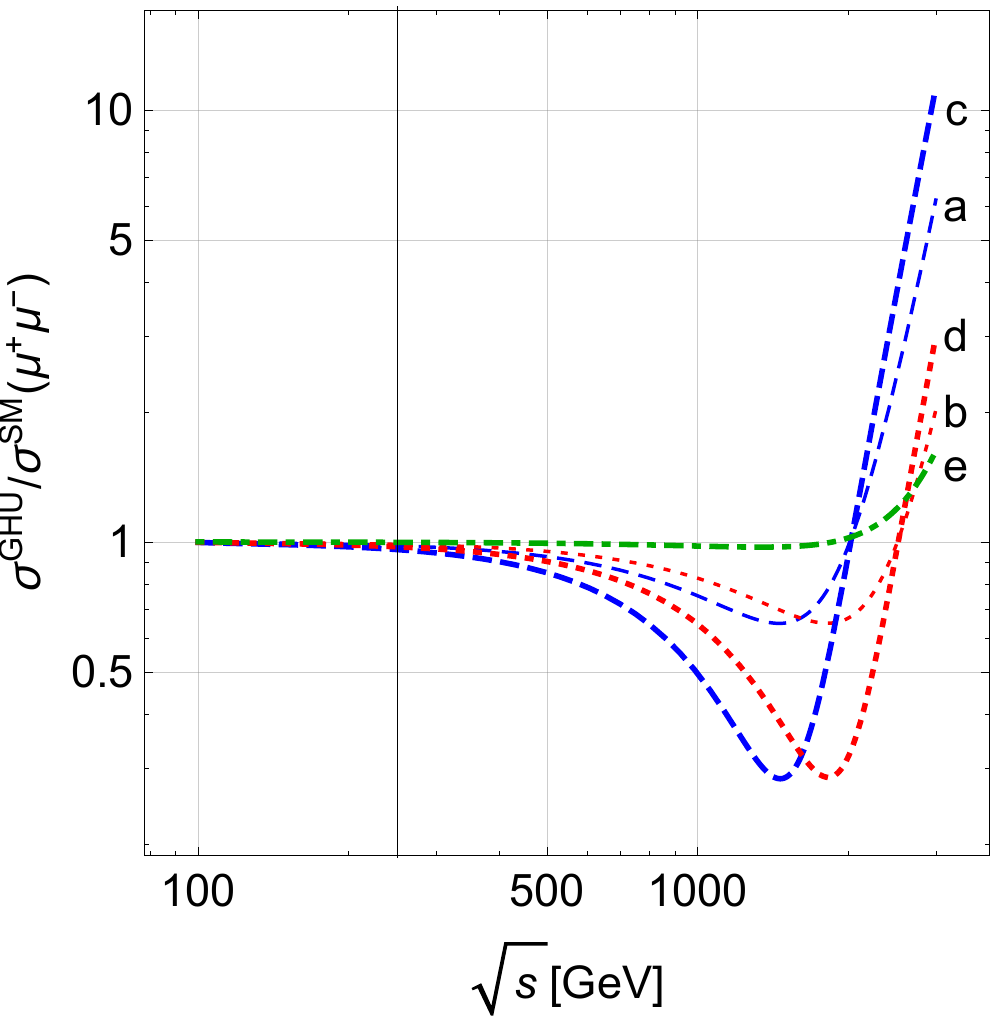}
\quad
\includegraphics[bb=0 0 360 232, width=7.cm]{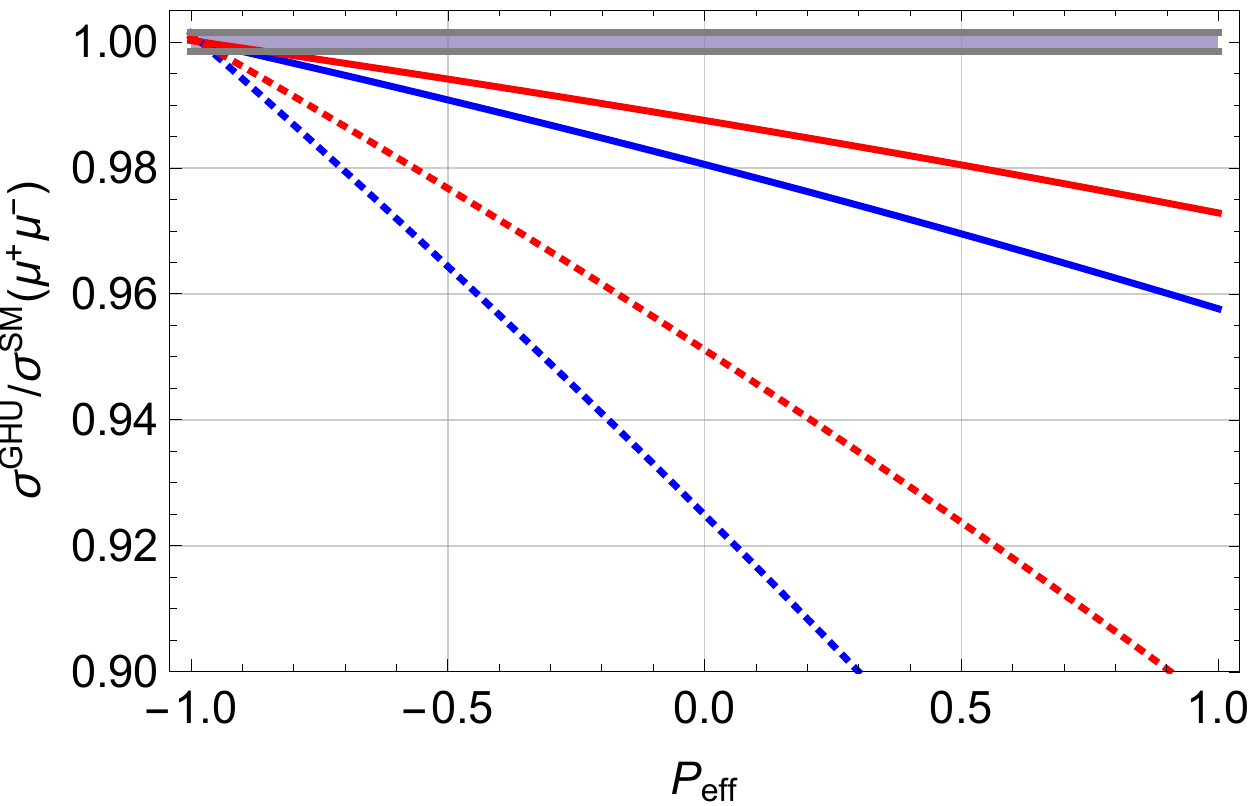}
\\
(a) \hskip 6cm (b)
\end{center}
\caption{
$F_1 = \sigma (\mu^+ \mu^- )^{\rm GHU}/\sigma (\mu^+ \mu^- )^\SM$ in (\ref{mupair1}) is plotted.
(a)  The $\sqrt{s}$ dependence is shown.
Blue curves a, c and green curve e are for $\theta_H = 0.0917$, whereas red curves b, d  are 
for $\theta_H = 0.0737$.  Curves a and b are with $P_\eff =0$.  Curves c and d are with $P_\eff =0.877$. 
Curve e is with $P_\eff =- 0.877$. 
(b) The polarization $P_\eff$ dependence is shown.
Solid (dashed) lines are for $\sqrt{s} = 250\,$GeV (500$\,$GeV).  
Blue lines are for $\theta_H = 0.0917$, whereas red lines  are for $\theta_H = 0.0737$.
The grey band indicates statistical uncertainty at $\sqrt{s} = 250\,$GeV with 
250$\,$fb$^{-1}$ data set.
}
\label{fig:mupair}
\end{figure}

The $\sqrt{s}$ dependence of $F_1$ is depicted in fig.~\ref{fig:mupair} (a).
The deviation from the SM becomes very large at $\sqrt{s} = 1.5\,{\rm TeV} \sim
2\,{\rm TeV}$ for $\theta_H = 0.09 \sim 0.07$, particularly with $P_\eff \sim 0.8$.
For $P_\eff \sim - 0.8$ the deviation is tiny.
At the energy $\sqrt{s} = 250\,$GeV the deviation might look small.  
As the  event number expected at ILC is so huge that deviation can be
unambiguously observed even at $\sqrt{s} = 250\,$GeV.
In fig.~\ref{fig:mupair} (b) the polarization $P_\eff$ dependence of $F_1$ is
depicted for $\sqrt{s} = 250\,$GeV and 500$\,$GeV.
As the polarization $P_\eff$ varies from $-1$ to $+1$, deviation from the SM
becomes significantly larger.  The grey band in fig.~\ref{fig:mupair} (b) indicates 
statistical uncertainty at $\sqrt{s} = 250\,$GeV with  250$\,$fb$^{-1}$ data set
in the SM.   It is seen that the signal of GHU can be clearly seen by measuring
the polarization dependence in the early stage of ILC 250$\,$GeV.

\subsection{Forward-backward asymmetry}

Not only in the total cross sections but also in differential cross sections for 
$e^+ e^- \go \mu^+ \mu^- $
significant deviation from the SM can be seen.\cite{Richard2018, Suehara2018}
Even with unpolarized beams the differential cross sections $d\sigma/d\cos\theta$
becomes 8\% (4\%) smaller than in the SM in the forward direction for $\theta_H = 0.0917$ ($0.0737$).

\begin{figure}[thb]
\begin{center}
\includegraphics[bb=0 0 360 243, width=6.7cm]{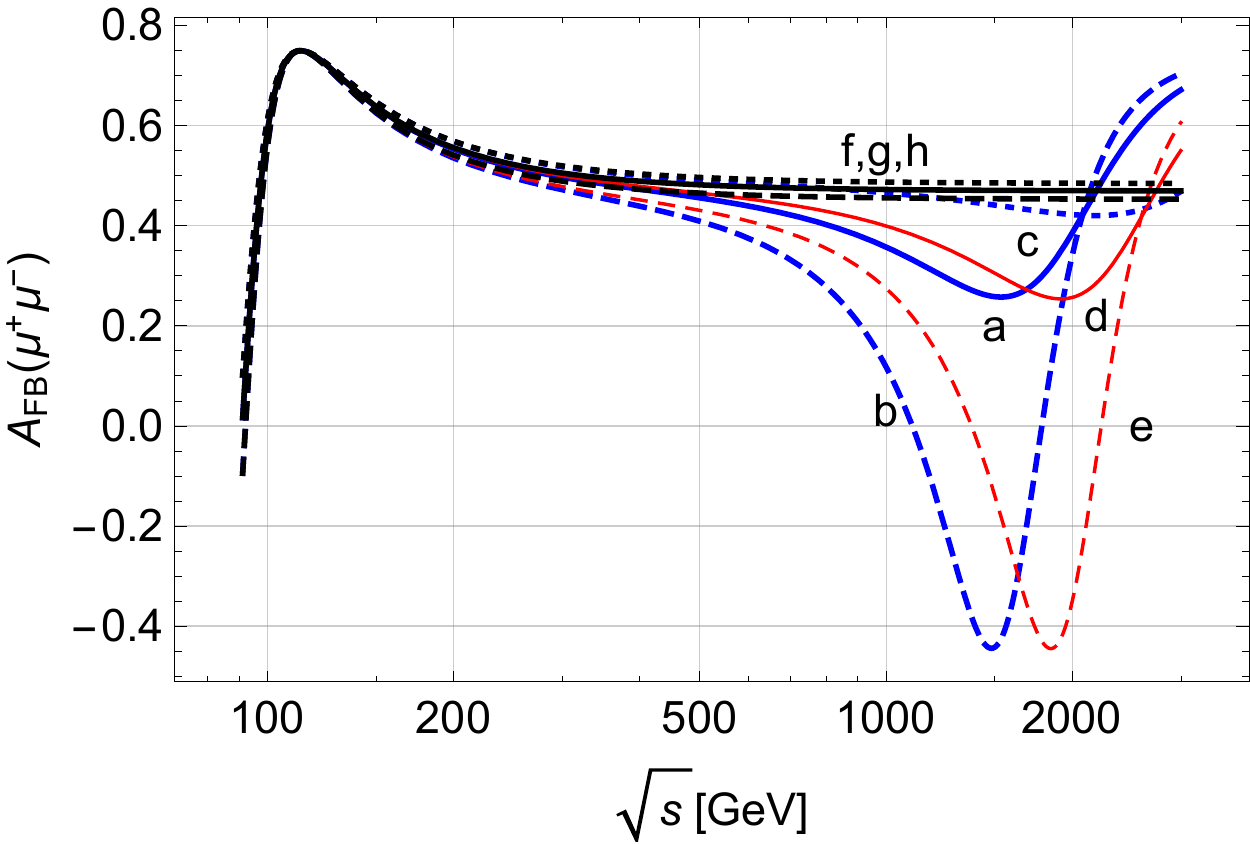}
\quad
\includegraphics[bb=0 0 360 230, width=6.8cm]{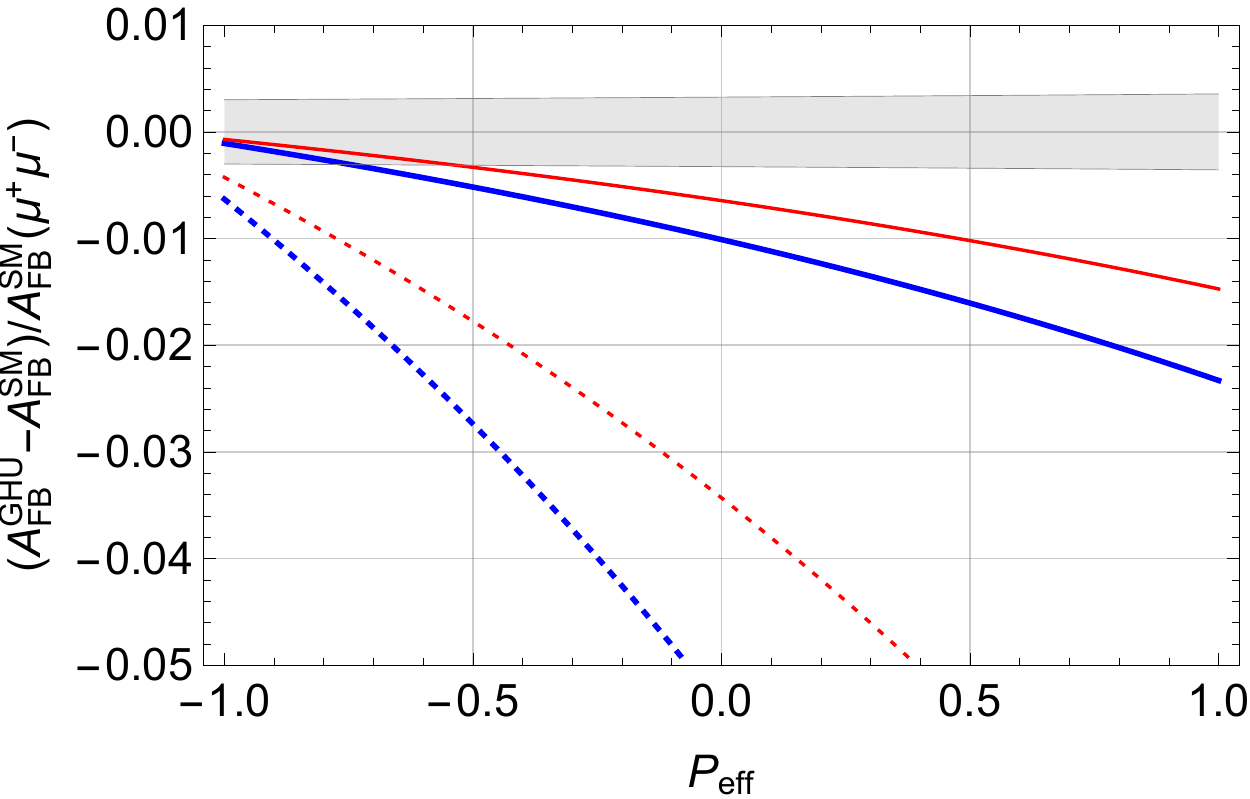}
\\
(a) \hskip 6.5cm (b)
\end{center}
\caption{
Forward-backward asymmetry $A_{\rm FB} (\mu^+ \mu^-)$.
(a)  The $\sqrt{s}$ dependence is shown.
Blue curves a, b, c are for $\theta_H = 0.0917$, red curves d, e are for $\theta_H = 0.0737$,
and black curves f, g, h are for the SM.
Solid curves a, d, f   are for unpolarized beams.
Dashed curves b, e, g  are with $P_\eff =0.877$.
Dotted curves c and h are with $P_\eff =- 0.877$.
(b) $(A_{\rm FB}^{\rm GHU} - A_{\rm FB}^{\rm SM})/A_{\rm FB}^{\rm SM}(\mu^+\mu^-)$ 
as functions of the effective polarization $P_\eff$. 
Solid and dotted lines are for $\sqrt{s} = 250\,$GeV and $500 \,$GeV, respectively.
Blue and red lines correspond to $\theta_H = 0.0917$ and $0.0737$, respectively.
The gray band indicates the  statistical uncertainty at $\sqrt{s}=250\,$GeV with $250\text{ fb}^{-1}$ data.
}
\label{fig:AFB}
\end{figure}

Forward-backward asymmetry $A_{\rm FB}$ characterizes this behavior.
In fig.~\ref{fig:AFB}(a) the $\sqrt{s}$-dependence of $A_{\rm FB}$ for $e^+ e^- \go \mu^+ \mu^- $
is shown.   As $\sqrt{s}$ increases the deviation from the SM becomes evident.
Again the deviation becomes largest around $\sqrt{s} = 1.5 \sim 2\,$TeV with $P_\eff = 0.877$
for $\theta_H = 0.0917 \sim 0.0737$. The sign of $A_{\rm FB}$ flips around $\sqrt{s} = 1.1 \sim 1.5\,$TeV.

Even at $\sqrt{s} = 250\,$GeV, significant deviation from the SM can be seen in the
dependence on the polarization ($P_\eff$) of the electron/positron beam as depicted in 
fig.~\ref{fig:AFB}(b).  With $250\text{ fb}^{-1}$ data the deviation amounts to 6$\sigma$ (4$\sigma$)
at $P_\eff = 0.8$ for $\theta_H = 0.0917 \, ( 0.0737)$, whereas
the deviation is within  an error at $P_\eff = - 0.8$. 
Observing the polarization dependence is a definitive way of investigating the details of the theory.

\subsection{Left-right asymmetry}

Systematic errors in the normalization of the cross sections are reduced in the  measurement of
\begin{align}
R_{f, LR} (\overline{P})
=\frac{\sigma( \bar{f}f \, ; \,  P_{e^-} = + \overline{P},  P_{e^+}=0 )}
{\sigma( \bar{f}f \, ; \, P_{e^-} = - \overline{P},  P_{e^+}=0 )} 
\label{defRfRL}
\end{align} 
where the electron beams are polarized with $P_{e^-} = + \overline{P}$  and $- \overline{P}$.
Only the polarization of the electron beams is flipped in experiments.  
Let $\sigma_{LR}^f$ ($\sigma_{RL}^f$) denote
the $e_L^-e_R^+ (e_R^- e_L^+) \to f\bar{f}$ scattering cross section.
Then the left-right asymmetry  $A_{LR}^f$ is related to $R_{f, LR} $ by
\begin{align}
A_{LR}^f &= 
\frac{\sigma_{LR}^f- \sigma_{RL}^f}{\sigma_{LR}^f + \sigma_{RL}^f}
= \frac{1}{\overline{P}}  \, \frac{1- R_{f,LR}}{1+ R_{f,LR}} ~.
\label{LRasym}
\end{align}
 
The predicted $R_{f, LR} (\overline{P})$ is summarized in Table \ref{tbl:LRasym}
for $\overline{P} = 0.8$.
Even at $\sqrt{s} = 250\,{\rm GeV}$ with $L_{int} = 250\,{\rm fb}^{-1}$ data, 
namely in the early stage of the  ILC experiment, 
significant deviation from the SM is seen. 
The difference between $R_{\mu, LR}$ and $R_{b, LR}$ stems from the different behavior of  
wave functions of $\mu$ and $b$  in the fifth dimension.

\begin{table}[htbp]
\caption{$R_{f, LR} (\overline{P})$  in the SM, and
deviations of $R_{f, LR} (\overline{P})^{\rm GHU} / R_{f, LR} (\overline{P})^{\rm SM}$
from unity are tabulated for $\overline{P} = 0.8$.
Statistical uncertainties of $R_{f,LR}^{\rm SM}$ is estimated with $L_{int}$ data for both 
$\sigma( \bar{f}f  ;   P_{e^-} = + \overline{P})$
and $\sigma( \bar{f}f  ;   P_{e^-} = - \overline{P})$, namely with $2 L_{int}$ data in all.}
\label{tbl:LRasym}
\vskip 8pt
\centering
\renewcommand{\arraystretch}{1.1}
\begin{tabular}{|c|c|c|cc|}
\hline
$f$ & $\sqrt{s}$~~~,~~~ $L_{int}$ & SM &\multicolumn{2}{c|}{ GHU} \\
&& $R_{f,LR}^{SM}$ (uncertainty) & $\theta_H=0.0917$ & $\theta_H = 0.0737$ \\
\hline
$\mu$ & $250\,{\rm GeV}$,  $250\,{\rm fb}^{-1}$ & $0.890$ ($0.3\%$) & $-3.4\%$ & $-2.2\%$ \\
      & $500\,{\rm GeV}$, $500\,{\rm fb}^{-1}$ & $0.900$ ($0.4\%$) & $-13.2\%$ & $-8.6\%$ \\
\hline
$b$   & $250\,{\rm GeV}$, $250\,{\rm fb}^{-1}$ & $0.349$ ($0.3\%$) & $-3.1\%$ &  $-2.1\%$ \\
      & $500\,{\rm GeV}$, $500\,{\rm fb}^{-1}$ & $0.340$ ($0.5\%$) & $-12.3\%$ & $-8.3\%$ \\     
\hline 
$t$   & $500\,{\rm GeV}$, $500\,{\rm fb}^{-1}$ & $0.544$ ($0.4\%$) & $-13.0\%$ & $-8.2\%$ \\
\hline
\end{tabular}
\end{table}

\section{Summary}
Gauge-Higgs unification predicts large parity violation in the quark-lepton couplings to
the $Z'$ bosons ($Z^{(1)}, \gamma^{(1)},Z_R^{(1)}$).  Although these $Z'$ bosons are
very heavy with masses 7 - 8$\,$TeV, they give rise to significant interference effects
in $e^+ e^-$ collisions at $\sqrt{s} = 250\,{\rm GeV} \sim 1\,$TeV.
We examined the A-model of $SO(5) \times U(1) \times SU(3)$ gauge-Higgs unification,
and found that significant deviation can be seen at 250$\,$GeV ILC with 250$\,{\rm fb}^{-1}$ data.
Polarized electron and positron beams are indispensable.
All of the total cross section, differential cross section, forward-backward asymmetry, 
and left-right asymmetry for $e^+ e^- \go f \bar f$ processes show distinct dependence 
on the energy and polarization.

We stress that new particles of masses 7 - 8$\,$TeV can be explored at 250$\,$GeV ILC
by seeing the interference effect, but not by direct production.
This is possible at $e^+ e^-$ colliders because the number of $e^+ e^- \go f \bar f$ events
is huge.  Although the probability of  directly producing $Z'$ bosons is suppressed
by a factor $(s/m_{Z'}^2)^2$,  the interference term is suppressed only by a factor
of $s/m_{Z'}^2$.  This gives a big advantage over $p p$ colliders such as LHC.

In this talk the predictions coming from the A-model are presented.  
It is curious to see how predictions change in the B-model.
Preliminary study indicates the pattern of the polarization dependence is reversed
in the B-model in comparison with the A-model.
The B-model is motivated by the idea of grand unification, which, in my opinion, is
absolute necessity in GHU in the ultimate form.  The A-model cannot be
implemented in natural grand unification.
Satisfactory grand unification in GHU has not been achieved 
yet.\cite{HosotaniYamatsu2015, Furui2016, HosotaniYamatsu2017}, 
\cite{Burdman2003}-\cite{MaruYatagai2019}

There are many other issues to be solved in GHU.  
Mixing in the flavor sector, behavior at finite temperature, inflation in cosmology,
and baryon number generation are among them.  
I would like to come back to these issues in due course.

\section*{Acknowledgement}

This work was supported in part  by Japan Society for the Promotion of Science, 
 Grants-in-Aid  for Scientific Research,  No.\ 15K05052  and No.\ 19K03873.

\def\jnl#1#2#3#4{{#1}{\bf #2},  #3 (#4)}

\def\PTP{{\em Prog.\ Theoret.\ Phys.\  }}
\def\PTEP{{\em Prog.\ Theoret.\ Exp.\  Phys.\  }}
\def\NPB{{\em Nucl.\ Phys.} B}
\def\PLB{{\it Phys.\ Lett.} B}
\def\PRL{\em Phys.\ Rev.\ Lett. }
\def\PRD{{\em Phys.\ Rev.} D}
\def\AP{{\em Ann.\ Phys.\ (N.Y.)} }
\def\MPLA{{\em Mod.\ Phys.\ Lett.} A}
\def\IJMPA{{\em Int.\ J.\ Mod.\ Phys.} A}
\def\IJMPB{{\em Int.\ J.\ Mod.\ Phys.} B}
\def\PR{{\em Phys.\ Rev.} }
\def\JHEP{{\em JHEP} }
\def\JCAP{{\em JCAP} }
\def\JPA{{\em J.\  Phys.} A}
\def\JPG{{\em J.\  Phys.} G}
\def\ibid{{\em ibid.} }

\renewenvironment{thebibliography}[1]
         {\begin{list}{[$\,$\arabic{enumi}$\,$]}  
         {\usecounter{enumi}\setlength{\parsep}{0pt}
          \setlength{\itemsep}{0pt}  \renewcommand{\baselinestretch}{1.2}
          \settowidth
         {\labelwidth}{#1 ~ ~}\sloppy}}{\end{list}}

\end{document}